\documentclass[10pt,letterpaper,twocolumn]{article}
\usepackage[letterpaper,breaklinks,hidelinks]{hyperref}

\usepackage[10pt,nocopyright]{sigmin}
\usepackage[square,comma,numbers,sort&compress]{natbib}

\usepackage[T1]{fontenc}
\usepackage{amssymb}
\usepackage{mathrsfs}

\usepackage{times}
\usepackage{textcomp}
\usepackage{verbatim}
\usepackage{amsmath}
\usepackage{amsfonts}
\usepackage{xspace}
\usepackage{xcolor}
\usepackage{graphicx}
\usepackage{url}
\usepackage{listings}
\usepackage{balance}
\usepackage{gnuplot-lua-tikz}
\usepackage{subfig}
\usepackage{booktabs}
\usepackage{multirow}
\usepackage{rotating}
\usepackage{fancyvrb}
\usepackage{lastpage}
\usepackage{alltt}
\usepackage{etoolbox}
\usepackage{ifdraft}
\usepackage{cleveref} 
\usepackage{fancyhdr}

\usepackage{xxxnotes}

\usepackage{colorhist}
\usepackage{scp-macros} \usetikzlibrary{shapes.geometric}

\input{glyphtounicode}
\usepackage[noeka]{mathrmletter}

\DeclareMathAlphabet{\mathcal}{OMS}{cmsy}{m}{n}

\newcommand{\splitcaption}[2]{%
  \caption[\protect\sc@oneperiod #1.]{#1 #2}}
\makeatother

\crefname{figure}{Figure}{Figures}
\crefname{mysubfigure}{Figure}{Figures}
\Crefname{mysubfigure}{Figure}{Figures}
\newcommand{\thiscref}[1]{this \lcnamecref{#1}}
\newcommand{\Thiscref}[1]{This \lcnamecref{#1}}

\newcommand{\paper}{paper\xspace}

\newcommand{\SIM}{SIM\xspace}

\begin{document}

\input{code/fmt}

\makeatletter
\def\code{\protect\@code}
\newlength\code@factor
\setlength\code@factor{1pt}
\def\codefont{\fontsize{\f@size\code@factor}{\f@size\code@factor}\bfseries\rmfamily}
\begingroup
\catcode`_=\active
\gdef\code@activate@us{%
  \let_\textunderscore
  \catcode`\_=\active
}
\endgroup
\def\@code#1{\ifmmode\text{\@@code{#1}}\else\@@code{#1}\fi}
\def\@@code#1{\begingroup\code@activate@us
  \codefont
  \scantokens{#1\endinput}%
  \endgroup
}
\makeatother
\let\syscall=\code
\let\gpcode=\code
\let\cmd=\code
\let\gpcmd=\gpcode

\def\UrlFont{\codefont}

\title{A Revised and Verified Proof of the Scalable Commutativity Rule
  }
\author{
  Lillian Tsai\rlap{,}$^\dag$
  Eddie Kohler\rlap{,}$^\star$
  M. Frans Kaashoek\rlap{,}$^\dag$ and
  Nickolai Zeldovich$^\dag$ \\
  $^\dag$ MIT CSAIL \qquad $^\star$ Harvard University
  }
\date{}

\maketitle

\section{Introduction}
\label{sec:intro}

This \paper explains a flaw in the published proof of the Scalable Commutativity Rule (SCR)~\cite{SCR}, presents a revised and formally verified proof of the SCR in the Coq proof assistant, and discusses the insights and open questions raised from our experience proving the SCR.

\section{The Scalable Commutativity Rule}
\label{sec:rule}

In order to explore the connection between commutativity and scalability in practical systems, Clements et al.~\cite{SCR} defined a new type of commutativity called \emph{\SIM commutativity},\footnote{SIM stands for State-dependent, Interface-based, and Monotonic.} a property that can hold of certain interface specifications. This was used to state and prove the Scalable Commutativity Rule (SCR), which claims that every SIM-commutative interface has a conflict-free implementation---that is, on modern machines, a \emph{scalable} implementation. Even
if an interface is commutative only in a restricted context, there exists an
implementation that scales in that context.


The rest of \thiscref{sec:rule} summarizes SIM commutativity and the precise statement of the rule, and describes the published proof of the rule.

\subsection{Specifications}

Specifications are represented using \emph{actions}, where each action is either an
\emph{invocation} (representing an operation call with arguments) or a
\emph{response} (representing a return value).
Each invocation is made by a specific thread, and the corresponding response
is returned to the same thread.
The division into invocations and responses models blocking interfaces and concurrent operations~\cite{herlihy:linearizability}.
Invocations are written as ``\inv{\alpha}{$f$(args)}'' and responses are written as
``\res{\alpha}{retval},'' where overbars mark responses and Greek letters
represent thread IDs.

\def\historyexample{%
\inv{\alpha}{A}
\inv{\gamma}{B}
\inv{\beta}{C}
\res{\alpha}{A}
\res{\beta}{C}
\res{\gamma}{B}
\inv{\alpha}{D}
\res{\alpha}{D}
\inv{\beta}{E}
\inv{\gamma}{F}
\inv{\alpha}{G}
\res{\beta}{E}
\res{\alpha}{G}
\res{\gamma}{F}}

\def\historyexamplealpha{%
\inv{\alpha}{A}
\res{\alpha}{A}
\inv{\alpha}{D}
\res{\alpha}{D}
\inv{\alpha}{G}
\res{\alpha}{G}}

A particular execution of a system is a \emph{history} or \emph{trace}, which
is just a sequence of actions. For example,
\begin{small}
\[H = [\text{\scpline{\historyexample}}]\]
\end{small}%
consists of seven invocations and seven corresponding responses across
three different threads.
In a \emph{well-formed} history, each thread's actions alternate invocations
and responses, so each thread has at most one outstanding invocation at any
point. $H$ above is well-formed; for instance, in the thread-restricted
subhistory $\HRESTRICT{H}{\thread{$\alpha$}{$t$}} = $$[\scpline{\historyexamplealpha}]$, which selects
\(\thread{$\alpha$}{$t$}\)'s actions from $H$, invocations and responses alternate as required.

A \emph{specification} models an interface's behavior as a prefix-closed set of well-formed histories. A system execution is ``correct'' according to the specification if its trace is
included in the specification. For instance, if $\SPEC$ corresponded to the
POSIX specification, then $[\inv{\alpha}{\code{getpid}}, \res{\alpha}{\code{92}}] \in \SPEC$
(a process may have PID 92) but $[\inv{\alpha}{\code{getpid}},
\res{\alpha}{\code{ENOENT}}] \not\in \SPEC$ (the \code{getpid()} system call may
not return that error). A specification constrains both invocations
and responses: $[\inv{\alpha}{\code{NtAddAtom}}]$ is not in the POSIX
specification because \code{NtAddAtom} is not a POSIX system call.

An \emph{implementation} is an abstract machine that takes invocations and
calculates responses. The original proof of the SCR by Clements et
al.~\cite{SCR} (also presented in \Cref{sec:proof})
uses a class of machines on which conflict-freedom is
defined; a good analogy is a Turing-type machine with a
random-access tape, where conflict-freedom follows if the machine's operations
on behalf of different threads access disjoint portions of the tape.

An implementation $M$ \emph{exhibits} a history $H$ if, when fed $H$'s
invocations at the appropriate times, $M$ can produce $H$'s responses (so that
its external behavior equals $H$ overall). An implementation $M$ is
\emph{correct} for a specification $\SPEC$ if $M$'s responses always obey the
specification. This means that every history exhibited by $M$ is either in
$\SPEC$ or contains some invalid invocation.

\subsection{Commutativity}
\label{sec:rule:sim-commutativity}
\label{sec:topic:strong-commutativity}
\SIM commutativity aims to capture state dependence at
the interface level.
State dependence means \SIM commutativity must capture when
operations commute in some states, even if those same operations do
not commute in other states. \SIM commutativity captures this contextually,
without reference to any particular implementation's state: 
to reason about \emph{possible} implementations, \SIM commutativity captures the
scalability inherent in the interface itself.
This in turn makes it
possible to use the SCR early in software development,
during interface design.

Commutativity states that actions may be reordered without affecting eventual
results.
A history $H'$ is a \emph{reordering} of $H$ when $\HRESTRICT{H}{t} =
\HRESTRICT{H'}{t}$ for every thread $t$. This allows actions to be reordered
across threads, but not within them.
%
For example, if $H =
[\text{\scpline{\inv{\alpha}{A} \inv{\beta}{B} \res{\alpha}{A}}}
\allowbreak
\text{\scpline{\inv{\alpha}{C} \res{\beta}{B}\res{\alpha}{C}}}]$, then
$[\text{\scpline{\inv{\beta}{B} \res{\beta}{B} \inv{\alpha}{A} \res{\alpha}{A} \inv{\alpha}{C}
    \res{\alpha}{C}}}]$
is a reordering of $H$, but
$[\text{\scpline{\inv{\beta}{B} \inv{\alpha}{C} \res{\beta}{B} \res{\alpha}{C}
    \inv{\alpha}{A} \res{\alpha}{A}}}]$
is not, since it doesn't respect the order of actions in $\HRESTRICT{H}{\alpha}$.

Now, consider a history $H = \HX \HCONCAT \HY$ (where $\HCONCAT$
concatenates action sequences).
$\HY$ \emph{\SI-commutes} in $H$ when given any reordering
$\HY'$ of $\HY$, and any action sequence $\HSUFF$,
\[ \HX \HCONCAT \HY \HCONCAT \HSUFF \in \SPEC \text{~~~if and only if~~~} \HX \HCONCAT \HY' \HCONCAT \HSUFF \in \SPEC. \]
This definition captures state dependence at the interface level.  The action
sequence $\HX$ puts the system into a specific state, without specifying a
representation of that state (which would depend on an implementation).
Switching regions $\HY$ and $\HY'$
requires that the exact responses in
$\HY$ remain valid according to the specification even if $\HY$ is
reordered.
The presence of region $\HSUFF$ in both histories requires that reorderings of
actions in region $\HY$ cannot be distinguished by future operations, which is
an interface-based way of saying that $\HY$ and $\HY'$ leave the system in the
same state.

Unfortunately, \SI commutativity is not sufficient to prove the 
SCR.
To avoid certain degenerate cases,\footnote{%
  Consider an \code{undefinedbehavior} invocation
  that allows later invocations to return any value whatsoever. This
  resembles the specification governing C compilers. Many sequences containing
  \code{undefinedbehavior} SI-commute, even if the same sequences without
  \code{undefinedbehavior} require a non-scalable implementation; for
  instance, groups of sequentially-consistent \code{set} operations don't
  SI-commute, since future \code{get}s must return the latest value, but
  \code{set}s plus \code{undefinedbehavior} do SI-commute, since future
  \code{get}s are unconstrained. Most practical implementations cannot see the
  future, and so cannot choose unsafe scalable designs in the hope that
  \code{undefinedbehavior} will eventually occur.%
} %
the definition of commutativity must be
strengthened to be {\em monotonic} (the M in \SIM).
An action sequence $\HY$ \textbf{\SIM-commutes} in a history $H = \HX \HCONCAT
\HY$ when for any \emph{prefix} $P$ of any reordering of $\HY$ (including
$P = \HY$), $P$ \SI-commutes in $\HX \HCONCAT P$.
Equivalently, $\HY$ \SIM-commutes in $H$ when, given any prefix $P$ of
any reordering of $\HY$, any reordering $P'$ of $P$, and any action sequence
$\HSUFF$,
\[ \HX \HCONCAT P \HCONCAT \HSUFF \in \SPEC \text{~~~if and only if~~~} \HX \HCONCAT P' \HCONCAT \HSUFF \in \SPEC. \]

Like \SI commutativity, \SIM\ commutativity captures 
state dependence and interface basis.  Unlike \SI commutativity, \SIM
commutativity excludes cases where the commutativity of a region
changes depending on future operations.
The SCR relies on {\SIM} commutativity.


\subsection{Rule}

The Scalable Commutativity Rule formally states the following:
\begin{quotation}
\noindent Assume an interface specification $\SPEC$ that has a correct
implementation, and a history $H = \HX \HCONCAT \HY$
exhibited by that implementation.
Then whenever $\HY$ \SIM-commutes in $H$, there exists a correct
implementation of $\SPEC$ whose steps in $\HY$ are conflict-free.
Since, given reasonable workload assumptions, conflict-free operations
empirically scale on modern multicore hardware, this
implementation is scalable in $\HY$.
\end{quotation}

\subsection{Proof}
\label{sec:proof}
\begin{figure}
\begin{obeylines}
\begin{obeyspaces}
    \LN{~1}\(\RULEIMP(s, i) \equiv\)
    \LN{~2}~~\(t \gets \THREAD(i)\)
    \LN{~3}~~If \(\HEAD(s.h[t]) = \COMMUTE\):  \COMMENT{enter conflict-free mode}
    \LN{~4}~~    \(s.\V{commute}[t] \gets \TRUE\);~ \(s.h[t] \gets \TAIL(s.h[t])\)
    \LN{~5}~~If \(\HEAD(s.h[t]) = i\):
    \LN{~6}~~    \(s.h[t].\POP()\)
    \LN{~7}~~If \(\HEAD(s.h[t])\) is a response and \(\THREAD(\HEAD(s.h[t])) = t\):
    \LN{~8}~~    \(r \gets \HEAD(s.h[t])\)   \COMMENT{replay $s.h$}
    \LN{~9}~~else if \(s.h[t] \neq \EMULATE\):   \COMMENT{$H$ complete/input diverged}
    \LN{10}~~    \(H' \gets \text{a witness consistent with \(s.h[t]\)}\)
    \LN{11}~~    For each invocation \(x\) in \(H'\):
    \LN{12}~~        \(\tup{s.\V{refstate}, \_, \_} \gets \REFIMP(s.\V{refstate}, x)\)
    \LN{13}~~    \(s.h[u] \gets \EMULATE\) for each thread \(u\)
    \LN{14}~~If \(s.h[t] = \EMULATE\):
    \LN{15}~~    \(\tup{s.\V{refstate}, r, \_} \gets \REFIMP(s.\V{refstate}, i)\)
    \LN{16}~~else if \(s.\V{commute}[t]\):   \COMMENT{conflict-free mode}
    \LN{17}~~    \(s.h[t] \gets \text{tail}(s.h[t])\)
    \LN{18}~~else:   \COMMENT{replay mode}
    \LN{19}~~    \(s.h[u] \gets \text{tail}(s.h[u])\) for each thread \(u\)
    \LN{20}~~Return \(\tup{s, r}\)
\end{obeyspaces}
\end{obeylines}
\rule{\columnwidth}{0.5pt}
\caption{Constructed scalable implementation $\RULEIMP$ for history $H$ and
  reference implementation $\REFIMP$.}
\label{fig:proof}
\end{figure}

The published proof of the SCR proceeds by construction. We briefly describe how the proof proceeds (eliding certain details about interruptibility and thread switches).

Given a specification $\SPEC$, an arbitrary \emph{reference} implementation $\REFIMP$ satisfying $\SPEC$, and a specific history $H = X \HCONCAT Y$ generated by $\REFIMP$ where $Y$ SIM-commutes in $H$, the proof aims to construct an implementation $\RULEIMP$ that scales (is conflict-free) within the SIM-commutative region $Y$ of $H$, but is correct (generates responses compatible with $\REFIMP$) for any history.

$\RULEIMP$ operates in three modes: \emph{replay}, \emph{conflict-free}, and \emph{emulation} modes. Its state consists of three parts:

\begin{enumerate}
\item $s.h[t]$, a per-thread history initialized as $X\HCONCAT [\COMMUTE]\HCONCAT (Y|t)$ (where $\COMMUTE$ is a special marker);
\item $s.\V{commute}[t]$, a per-thread flag which is set to \TRUE\ when $\COMMUTE$ is reached;
\item $s.\V{refstate}$, the state of reference implementation $M$.
\end{enumerate}

\noindent \Cref{fig:proof} shows pseudocode for $\RULEIMP$'s implementation.

$\RULEIMP$ starts in replay mode. This mode replays the history $H$ as long as each thread invokes operations in the same order as it did in $H$.
While thread $t$ has not yet reached the commutative region, $t$'s invocation matches that of $s.h[t]$ (line 5), and the next action in $s.h[t]$ is a response to $t$'s invocation (line 7), $\RULEIMP$ returns the matching response (line 8) and advances the stored history $s.h[u]$ of all threads (line 19). 

If the next step in $s.h[t]$ is $\COMMUTE$, then $s.commute[t]$ is set to \TRUE\ (line 4) and $\RULEIMP$ enters conflict-free mode. In this mode, $\RULEIMP$ starts replaying steps in $Y|t$. Just like before, if the next action in $s.h[t]$ is a response to $t$'s invocation (line 7), $\RULEIMP$ returns the matching response (line 8). However, it advances only the stored history of $t$, namely $s.h[t]$ (line 17). 
This means that after $\RULEIMP$ enters conflict-free mode for a thread $t$ (i.e., $s.commute[t] = \TRUE$), it accesses and modifies only those state components specific to $t$, and any steps in the conflict-free region $Y$ are conflict-free.

Once $s.h[t]$ has fully replayed or if a thread $t$ \emph{diverges}---that is, $t$'s next invocation does not match the next invocation in $s.h[t]$---then $\RULEIMP$ switches to emulate mode for all threads. In this mode, $\RULEIMP$ feeds the reference machine $M$ invocations to determine the response to return. Before we can use the reference machine $M$ in this way, however, the state of $M$ must reflect the state of the execution history thus far.

We initialize $\REFIMP$'s state by finding a \emph{witness} of $\REFIMP$ that is consistent with the execution history (line 10). A \emph{witness} of the execution history is a sequence of invocations that, when fed to $\REFIMP$, generates the execution history.
Once a witness is found, we know that feeding $\REFIMP$ the witness's sequence of invocations (line 12) will bring $\REFIMP$ to a valid state, where all future responses will be valid according to the spec.

Finding a witness is easy before $\RULEIMP$ reaches conflict-free mode: since $\RULEIMP$ generated the history $X\HCONCAT Y$, simply take all invocations in $X$ (up to the current point) in order as the witness.
However, if $\RULEIMP$ has entered the conflict-free mode and executed operations in the SIM-commutative region, the order in which operations were executed in this region may not equal the order in which operations were executed in $Y$. Here is where SIM commutativity comes in: we can reorder the operations in the commutative region of the execution history to achieve a witness. Because of SIM commutativity, \textbf{we can initialize $\REFIMP$ with a witness
comprised of commutative actions in a different order than that in which they were executed, and all future responses will still be valid according to the specification}.

\section{The flaw}
\label{sec:flaw}

This last, bold statement is where the proof goes wrong. 
To help build intuition, we now present a counterexample in which the proof-constructed machine fails: the construction cannot find a witness that will initialize $\REFIMP$ with a valid state.

\subsection{Specification}
Imagine a specification $\SPEC$ for opening and closing files with the following two operations:
\begin{enumerate}
    \item int $\open$(): returns a currently-unused file descriptor with value $> 0$;
    \item int $\close$(int \textit{fd}): returns $\OK$ on success, $\EBADFD$ if \textit{fd} was never opened, or $\ECLOSEDFD$ if \textit{fd} was previously opened, but is currently closed.
\end{enumerate}

To better understand $\SPEC$, we look at a couple of examples of valid and invalid histories. Let $\alpha$ and $\beta$ be thread IDs. 
The following histories would be valid:
\begin{align*}
    H_1 &= [\text{\scpline{\inv{\alpha}{\open}
    \res{\alpha}{\code{1}}
    \inv{\beta}{\open}
    \res{\beta}{\code{2}}
    \inv{\alpha}{\close(1)}
    \res{\alpha}{\OK}}}]\\
    H_2 &= [\text{\scpline{\inv{\alpha}{\open}
    \res{\alpha}{\code{1}}
    \inv{\beta}{\close(2)}
    \res{\beta}{\EBADFD}}}]\\
H_3 &= [\text{\scpline{\inv{\alpha}{\open}
    \res{\alpha}{\code{1}}
    \inv{\alpha}{\close(1)}
    \res{\alpha}{\OK}
    \inv{\alpha}{\close(1)}
    \res{\alpha}{\ECLOSEDFD}}}] 
\end{align*}
and the following histories would be invalid:
\begin{align*}
    H_1' =& [\text{\scpline{\inv{\alpha}{\open}
    \res{\alpha}{\code{1}}
    \inv{\beta}{\open}
    \res{\beta}{\code{1}}}}]\\
    & \textit{(returns used FD)}\\
    H_2' =& [\text{\scpline{\inv{\alpha}{\open}
    \res{\alpha}{\code{1}}
    \inv{\alpha}{\close(2)}
    \res{\alpha}{\OK}}}]\\
    & \textit{(should return \EBADFD)}\\
    H_3' =& [\text{\scpline{\inv{\alpha}{\open}
    \res{\alpha}{\code{1}}
    \inv{\alpha}{\close(1)}
    \res{\alpha}{\OK}
    \inv{\alpha}{\close(1)}
    \res{\alpha}{\OK}}}] \\
    & \textit{(should return \ECLOSEDFD)}
\end{align*}

Note that all sequences of $\open$ operations are SIM-commutative regions: reordering any number of $\open$ operations satisfies $\SPEC$, since the returned FDs are still unique and positive in value. 

\subsection{Reference implementation}
We now choose a simple reference implementation $\REFIMP$ that implements $\SPEC$. The implementation has two pieces of global state, namely a counter $\V{gfd}$ initialized as 0 and a \V{closed} set initially empty. The two operations are implemented as follows:
\begin{enumerate}
    \item int $\open$: increment $\V{gfd}$ and return the new value;
    \item int $\close$(int \V{fd}): if $0 < \V{fd} \leq \V{gfd}$ and $\V{fd} \not\in \V{closed}$, then return $\OK$ and add $\V{fd}$ to $\V{closed}$. Otherwise if $\V{fd} \in \V{closed}$, return $\ECLOSEDFD$, else return $\EBADFD$.
\end{enumerate}

$\REFIMP$ satisfies $\SPEC$: $\open$ returns only unused, positive file descriptors since the counter never (disregarding overflows) repeats values. $\close(\V{fd})$ returns $\OK$ if the file has been opened, since all files below the current value of $\V{gfd}$ must have been opened before, and $\ECLOSEDFD$ if the file has already been closed. Otherwise, $\V{fd}$ is invalid and $\REFIMP$ returns $\EBADFD$.

Although $\REFIMP$ satisfies $\SPEC$, it is \emph{not} scalable for SIM-commutative regions: all $\open$ and $\close$ operations access and write the shared, global counter and list.

We now have all the pieces to implement (and break) the proof construction from \autoref{sec:proof}.

\subsection{Incorrect construction}
We first choose a SIM-commutative region of $\SPEC$. Since regions of $\open$ operations are SIM-commutative, we generate the following history using $\REFIMP$:
    $$H_{\text{commute}} = 
    [\text{\scpline{\inv{\alpha}{\open}
    \res{\alpha}{\code{1}}
    \inv{\beta}{\open}
    \res{\beta}{\code{2}}}}]
    $$
This history is used to set the state of $\RULEIMP$, initializing $s.h[*]$ as
\begin{align*}
s.h[\alpha] &= 
    [\text{\scpline{\inv{}{\COMMUTE}
    \inv{\alpha}{\open}
    \res{\alpha}{\code{1}}}}]\\
s.h[\beta] &= 
    [\text{\scpline{\inv{}{\COMMUTE}
    \inv{\beta}{\open}
    \res{\beta}{\code{2}}}}]
\end{align*}

Now we execute $\RULEIMP$ on the following sequence of operations: 
    $$\text{\scpline{\inv{\beta}{\open}
    \inv{\alpha}{\close(1)}}}$$

Following the proof construction, $\inv{\beta}{\open}$ will first cause $\RULEIMP$ to switch to conflict-free mode for $\beta$ (\Cref{fig:proof}, lines 3-4) because $s.h[\beta][0] = \COMMUTE$. After line 4, 
$$s.h[\beta] = 
    [\text{\scpline{\inv{\beta}{\open}
    \res{\beta}{\code{2}}}}]$$

Next, since the invocation $\open$ matches the first invocation by $\beta$ in $s.h[\beta]$, $\RULEIMP$ will return the value \code{2} (lines 5-8). From line 17, the current state is now 
\begin{align*}
s.h[\alpha] &= 
    [\text{\scpline{\inv{}{\COMMUTE}
    \inv{\alpha}{\open}
    \res{\alpha}{\code{1}}}}]\\
s.h[\beta] &= 
    [\text{\scpline{}}]
\end{align*}

The next invocation $\RULEIMP$ receives is $\inv{\alpha}{\close(1)}$. This switches $\RULEIMP$ to conflict-free mode for $\alpha$ (lines 3-4), and sets the state to
$$s.h[\alpha] = 
    [\text{\scpline{\inv{\alpha}{\open}
    \res{\alpha}{\code{1}}}}]$$

$\RULEIMP$ cannot, however, replay $s.h[\alpha]$ as it did for $\beta$. $\close(1)$ represents a divergence from $H_{\text{commute}}$, so $\RULEIMP$ enters emulate mode (lines 9-13).
Our proof construction will now have to get $\REFIMP$ to a valid state consistent with the current history (line 10) so that we can feed $\REFIMP$ future invocations during emulation phase (as shown in line 15).

At this point, our recorded history is $$\Hcurrent = [\text{\scpline{\inv{\beta}{\open}
    \res{\beta}{2}}}]$$
But even though this history is correct (it's in $\SPEC$), the reference
implementation \emph{cannot exhibit this history}. %
In $\REFIMP$, the first call to $\open$ always returns file descriptor 1.

This causes a serious problem. The construction must initialize
$s.\V{refstate}$ to some value that could have exhibited $\Hcurrent$, but no
such state exists. As a result, any initialization sequence attempted by $\RULEIMP$
can cause errors on future invocations.
%
%
Invoking $\open$ only once sets $\V{gfd} = 1$; a following $\open$ call would return $2$, an invalid response (since 2 is still open). Invoking $\open$ two or more times would avoid this problem, but would cause a following $\close$(1) call to return the incorrect error ($\ECLOSEDFD$ instead of $\EBADFD$). Thus, our proof construction fails both to find a witness consistent with the current execution history and to initialize $\REFIMP$ with a valid state.

\subsection{When might the proof fail?}
Note that $\REFIMP$ can reach a valid state if either \emph{none} or \emph{all} of the SIM-commutative region in question has been replayed. In other words, if $\RULEIMP$ diverges before or after the SIM-commutative region, then we can always get $\REFIMP$ to a valid state. 
For the former, when $\RULEIMP$ diverges before the SIM-commutative region, we can just feed $\REFIMP$ the recorded history's invocations sequentially in the order in which they occurred (that order is precisely known).
For the latter, when $\RULEIMP$ diverges after the SIM-commutative region, we know there is at least one ordering of \emph{all} operations in the SIM-commutative region that $\REFIMP$ can generate, namely $H_{\text{commute}}$. Thus, feeding $\REFIMP$ all possible orderings of all the operations in the SIM-commutative region until $\REFIMP$ reaches a valid state must eventually terminate.

Furthermore, if a prefix of the operations in the SIM-commutative region \emph{with the same order as in $H_{\text{commute}}$} occurs before divergence, then feeding $\REFIMP$ the operations in this order will also bring $\REFIMP$ to a valid state. This is because $\REFIMP$ generated $H_{\text{commute}}$ by being fed operations sequentially in this order.

The potential for failure arises only when a \emph{prefix of a reordering} of the SIM-commutative region occurs before divergence, as demonstrated in our example.
In this scenario, we do know not know if this reordered sequence of invocations and responses can ever be generated by $\REFIMP$.

\subsection{Why does the proof fail?}

The key problem is that our reference implementation $\REFIMP$ cannot exhibit some histories required for SIM commutativity. $\REFIMP$ is correct, so every history $H$ exhibited by $\REFIMP$
is in $\SPEC$, but there may be some histories in $\SPEC$ that $\REFIMP$ cannot exhibit.
That is, the reference implementation's \emph{exhibited specification}---the set of histories that it can possibly exhibit---may lack some traces in $\SPEC$.
When this gap arises, and regions that are SIM-commutative in $\SPEC$ do not commute in the exhibited specification, then $\REFIMP$ may not be able to achieve a state consistent with the history our construction requires, as shown in our example.

On the other hand, the proof construction likely works whenever the given region SIM-commutes in the \emph{exhibited} specification. Put another way, our construction likely works as long as
``$\REFIMP$, the reference implementation, produces the same results for any reordering of the commutative region'' (quoted from the published version of the proof in Clements et al.~\cite{SCR}). Any prefix of a reordering of the SIM-commutative region would still SIM-commute in the exhibited specification because $\REFIMP$ produces the same result for any ordering, and thus we can eliminate the failure case of our proof construction.

\subsection{How to fix the proof?}
We have seen that a bad reference implementation can prevent our proof technique from producing an implementation that scales within a given commutative region. We considered 
several fixes for this issue.

\begin{enumerate}
    \item \emph{Exhibited specification}. As we noted above, the proof construction will likely work if we restrict the rule to regions that are SIM-commutative in the exhibited specification, rather than those that are SIM-commutative in the defined specification. 
    \item \emph{Specification oracle}. Alternately, we could remove the reference implementation from the proof entirely, and instead rely on a specification oracle that enumerates valid responses to invocations.
\end{enumerate}

Exhibited specifications would preserve the somewhat ``practical" feel of the flawed proof, and the useful intuition that a scalable implementation can be obtained from a non-scalable implementation by logging and reconciliation. However, this is stricter than SIM commutativity, which places requirements on the specification, not the implementation.
This prevents the SCR from, for example, informing programmers about potential areas to increase the scalability of their implementations.

Specification oracles feel less realistic than reference implementations, but they have the advantage of completely avoiding the issue of whether a given specification can be implemented at all. They also fit nicely into Coq. Because of these reasons, our 
machine-verified proof uses specification oracles. 

\section{The Verified Proof} 
\label{sec:verified_proof}
\lstset{
  basicstyle=\footnotesize\ttfamily,
  columns=fullflexible,
  mathescape
}

\begin{figure}
\begin{obeylines}
\begin{obeyspaces}
    \LN{~1}\(\ORACLEIMP(s, i) \equiv\)
    \LN{~2}~\(t \gets \THREAD(i)\)
    \LN{~3}~\(h_{copy} \gets []\)
    \LN{~4}~\(h_{perf} \gets []\)
    \LN{~5}~If \(s.mode \neq \ORACLE\):
    \LN{~6}~~~~~If \(s.X\_copy = []\): \COMMENT{enter conflict-free mode}
    \LN{~7}~~~~~     \(s.mode \gets \CONFLICTFREE\)
    \LN{~8}~~~~~     \(h_{copy} \gets s.Y\_copy[t]\)
    \LN{~9}~~~~~     \(h_{perf} \gets s.Y\_performed[t]\)
    \LN{10}~~~~~else: \COMMENT{still in replay mode}
    \LN{11}~~~~~     \(h_{copy} \gets s.X\_copy\)
    \LN{12}~~~~~     \(h_{perf} \gets s.X\_performed\)
    \LN{13}~~~~~If \(\HEAD(h_{copy}) = i\):
    \LN{14}~~~~~     \(h_{copy}.\POP()\)
    \LN{15}~~~~~If \(\HEAD(h_{copy})\) is a response \& \(\THREAD(\HEAD(h_{copy})) = t\):
    \LN{16}~~~~~     \(r \gets \HEAD(h_{copy})\)
    \LN{17}~~~~~     \(h_{perf}.\APPEND((i, r))\)
    \LN{18}~~~~~else: \COMMENT{$h.{copy}$ empty or input diverged}
    \LN{19}~~~~~     \(s.mode \gets \ORACLE\)
    \LN{20}~If \(s.mode = \ORACLE\):
    \LN{21}~~   for each possible response $resp$ to invocation $i$:
    \LN{22}~~        \(H' \gets \text{a history consistent with performed actions}\)
    \LN{23}~~        If $O_\SPEC(H' \HCONCAT [(i,resp)]) = \TRUE$:
    \LN{24}~~             \(r \gets resp\)
    \LN{25}~~             \(s.oracle\_performed.\APPEND((i,r))\)
    \LN{26}~~             \text{break}
    \LN{27}~else if \(s.mode = \CONFLICTFREE\):
    \LN{28}~    \(s.Y\_copy[t] \gets \text{tail}(s.Y\_copy[t])\)
    \LN{29}~    \(s.Y\_performed[t] \gets h_{perf}\)
    \LN{30}~else:   \COMMENT{replay mode}
    \LN{31}~    \(s.X\_copy \gets \text{tail}(s.X\_copy)\)
    \LN{32}~    \(s.X\_performed \gets h_{perf}\)
    \LN{33}~Return \(\tup{s, r}\)
\end{obeyspaces}
\end{obeylines}
\rule{\columnwidth}{0.5pt}
\caption{Verified constructed scalable implementation $\ORACLEIMP$ for history $H$ and reference implementation $\REFIMP$.}
\label{fig:verified_proof}
\end{figure}

\Thiscref{sec:verified_proof} describes our machine-verified proof of the SCR based on a specification oracle.
The pseudocode for our proof construction $\ORACLEIMP$ is shown in~\Cref{fig:verified_proof}.

\subsection{Oracle proof construction}
An oracle $O_\SPEC$ is a function from a history $H$ to $\{\TRUE,\FALSE\}$ defined as 
\[
O_\SPEC(H) =
\begin{cases} \TRUE & \text{$H \in \SPEC$}\\
\FALSE & \text{$H \not\in \SPEC$}\\
\end{cases}\]

Given a specification $\SPEC$, an oracle $O_\SPEC$, and a specific history $H = X \HCONCAT Y$ where $Y$ SIM-commutes in $H$, the proof constructs an implementation $\ORACLEIMP$ that executes conflict-free within the SIM-commutative region $Y$.

$\ORACLEIMP$ operates in three modes: \emph{replay}, \emph{conflict-free}, and \emph{oracle} modes. Its state consists of three parts (with sub-parts):
\begin{enumerate}
\item Copies of $H$ as histories to replay:
\begin{itemize}
    \item $s.X\_copy$, a global list of actions initialized as $X$
    \item $s.Y\_copy[t]$, a per-thread list of actions initialized as $Y|t$
\end{itemize}

        Note that these copies of $H$ are equivalent to the per-thread copies $s.h[t]$ of $H$ used in the original proof construction. The split here into $X\_copy$ and $Y\_copy[t]$ allows us to cleanly separate the global state from the per-thread state in our Coq formulation. Instead of checking for a $\COMMUTE$ marker to indicate when the $Y$ region has been reached, our construction will check if $X\_copy$ is empty.

\item Lists of performed actions:
\begin{itemize}
    \item $s.X\_performed$, a global list of performed actions of $X$ initialized as []
    \item $s.Y\_performed[t]$, a per-thread list of performed actions of $Y|t$ initialized as [] 
    \item $s.oracle\_performed$, a global list of performed actions in oracle mode initialized as [] 
\end{itemize}
\item $s.mode$, a global flag indicating the current mode of the machine
\end{enumerate}

The replay and conflict-free modes act similar to the corresponding modes of $\RULEIMP$ from the prior proof. If $\ORACLEIMP$ is not already in oracle mode, then $\ORACLEIMP$ is in replay mode if $s.X\_copy$ is nonempty, or in conflict-free mode if $s.X\_copy$ is empty (line 6). If both $s.X\_copy$ and $s.Y\_copy[t]$ for all $t$ are empty, $\ORACLEIMP$ switches to oracle mode (line 19). 

In replay mode, if the next requested invocation matches the next invocation in $s.X\_copy$ and the next action in $s.X\_copy$ is a response to that invocation, $\ORACLEIMP$ pops the head off of $s.X\_copy$, returns the response, and appends the response to $s.X\_\allowbreak performed$ (lines 13-17). 
Otherwise, $\ORACLEIMP$ has diverged from $X \HCONCAT Y$  and switches to oracle mode.

In conflict-free mode (set up in lines 7-9), if the next requested invocation by $t$ matches the next invocation in $s.Y\_copy[t]$ and the next action in $s.Y\_copy[t]$ is a response to that invocation, $\ORACLEIMP$ returns the response and appends the response to $s.Y\_\allowbreak performed[t]$ (lines 13-17). Otherwise, the execution has diverged and $\ORACLEIMP$ switches to oracle mode.

In oracle mode, the next response is determined by querying the oracle function. $\ORACLEIMP$ iterates through all possible responses $r$ to the invocation $i$ and calls $O_\SPEC(H' \HCONCAT [r])$, 
where $H'$ is a history consistent with the performed actions in the history ($s.X\_performed$, $s.Y\_\allowbreak performed[t]$ for all $t$, and $s.oracle\_performed$). If the oracle returns $true$, then $\ORACLEIMP$ stops iterating, returns $r$, and the chosen response is appended to $s.oracle\_\allowbreak performed$ (lines 20-25).

More specifically, $H'$ is constructed as 
\begin{lstlisting}
    $s.X\_performed$
    $\HCONCAT s.Y\_performed[t_0] \HCONCAT \dots \HCONCAT s.Y\_performed[t_{\#threads}]$
    $\HCONCAT s.oracle\_performed$
\end{lstlisting}
Because of SIM commutativity, any ordering of operations in $Y$ satisfies the spec. Thus, sequentially concatenating the $s.Y\_performed[*]$ to construct $H'$ generates a valid history.

Note that while in conflict-free mode, $\ORACLEIMP$ executes in a scalable way: no thread accesses another's state. Thus, $\ORACLEIMP$ should satisfy the SCR. In the next section, we describe how we verified this claim.

\subsection{Coq formalization}
Here we give an overview of how we formalized the proof construction and proved its correctness in Coq. The complete Coq source is available at \url{https://github.com/tslilyai/coq_scr}.

\subsubsection{Definitions}
Our Coq model includes definitions for action histories (see \Cref{sec:rule}), the machine state and modes as described above, conflict-freedom, SIM commutativity, and machine execution. These definitions are presented in the Appendix (\Cref{fig:scrdefns}). Actions are tuples of 
$\tup{threadID, op, response}$, and histories are (reversed) lists of actions. We create an enum for modes, and use a record to encode state, where the record contains either thread-specific or global histories. Per-thread state is represented as a function from tid to history.

Defining conflict-freedom requires defining conflict-free writes and reads.
To define conflict-free writes, we use two constructions: \code{diff\_histories\_tid\_set} takes two histories and returns the set of threads whose per-thread histories ($s.Y\_performed[t]$ or $s.Y\_copy[t]$) change from the first history to the second. \code{diff\_states\_tid\_set} takes two states and uses \code{diff\_histories\_tid\_set} to return the set of threads that have had their per-thread state changed between the two states. Note that per-thread state in our construction changes only if per-thread history changes.

With these constructions, we can define conflict-free writes (\code{conflict\_free\_writes}). All writes performed during a step of the machine from state $s_1$ to $s_2$ on thread $t$ are conflict-free writes if:
\begin{itemize}
    \item the global state of $s_1$ and $s_2$ have equivalent values
    \item either the per-thread state of $s_1$ and $s_2$ have equivalent values, or the \code{diff\_states\_tid\_set}$(s_1, s_2)$ must contain at most the single calling thread $t$
\end{itemize}
These restrictions guarantee that any writes made during this step only modify $t$'s state. 

We define conflict-free reads (\code{conflict\_free\_reads}) as follows: say the machine takes a step $i$ on thread $t$ from state $s_1$. Then this step performs only conflict-free reads if, for any state $s_2$ with the same per-thread state of $t$ and the same mode as $s_1$,
\begin{itemize}
    \item the machine returns the same value when executing $i$ from $s_2$ as it does executing $i$ from $s_1$
    \item the state of the mode is unchanged when the machine executes $i$ from either $s_1$ or $s_2$
\end{itemize}
Equivalent return values ensure that no read during the machine step depends on another thread's state or any global state apart from the mode; $s_1$ and $s_2$ are only restricted in their values of the per-thread state of $t$ and the mode.
Unchanging modes ensure that the only piece of global state (the mode) read by the machine in executing the step will not be modified during the step, which would cause the read to conflict with the modifying write. 

With both conflict-free writes and conflict-free reads, we can prove a step is conflict-free by proving that the step's writes and reads are both conflict-free.

Note that this definition of conflict-freedom is quite different than the one presented in by Clements et al. in the original SCR paper, which reasoned about conflicts in terms of memory access sets. Instead, our Coq definition is specialized for the mechanics of our proof construction, and allows us to reason on the much higher level of our construction's abstract per-thread state (e.g., $s.Y\_copy[t]$) rather than individual memory accesses.

\subsubsection{Theorem statements}
The final theorem and important lemmas are shown in the Appendix (\Cref{fig:scrthms}).
Key helpers to prove these lemmas include determining the current mode and state of the machine, definitions for switching between modes when appropriate, and lemmas proving correctness of the machine when the machine is at each mode.

Our proof strategy was to first prove two lemmas, namely \code{machine\_correct}, which states that the machine $\ORACLEIMP$ generates only histories satisfying the spec, and \code{machine\_conflict\_free}, which states that the machine execution is conflict-free during the SIM-commutative region $Y$. We used these lemmas to prove our final statement of the SCR:
\begin{enumerate}
    \item Correctness: All histories achievable by $\ORACLEIMP$ satisfy the spec and $\ORACLEIMP$ never returns an invalid response
    \item Conflict-freedom: Any step $\ORACLEIMP$ takes in the SIM-commutative region $Y$ is conflict-free
\end{enumerate}

\subsection{Proof assumptions and evaluation}
Our Coq proof makes several assumptions, encoded as parameters or as part of the definition:
\begin{itemize}
    \item The oracle can enumerate all possible responses. For our proof, we assume something stronger, namely that the number of responses is finite; finiteness was necessary to convince Coq that the oracle function terminates.
    \item For every invocation that is valid to call, there exists a valid response to return. 
    \item The oracle is correct: for all $H$, $O_\SPEC(H) = \TRUE \iff H \in \SPEC$.
\end{itemize}

We made a number of decisions to ease the proof process. As we described earlier, we reason about conflict-freedom at a high level (without, for example, modeling memory arrays or low-level memory accesses as we did in our first attempt). This abstraction greatly simplified the proof process. We also switched to local reasoning, i.e., reasoning about steps of the machine, rather than proving facts about the machine's entire history, and we found that non-inductive definitions (for example, for reordered histories) made the proofs easier to handle.
Finally, we proved all lemmas using the reverse of histories: because histories are defined as a list, the generated inductive cases were more intuitive (new actions are added to the head, not the tail, of the list). 

The entire Coq development is 2056 LOC, of which 562 lines are state definitions, lemmas, and theorems.
The effort took approximately 3 person-months, including several weeks stuck on verifying the original (incorrect) proof. 
We believe that without attempting to verify the published proof in Coq, the flaw in the proof may have been difficult to find.

\section{Discussion and Conclusion}
\label{sec:disc}

We have presented an initial proof for the SCR, an example and brief discussion about how it is incorrect, and a new, verified proof for the SCR.
However, the new proof is not wholly satisfactory. Most notably, the verified proof relies on the existence of a specification oracle, which causes it to stray even further from an imaginable construction than the original proof. We also assume enumerable responses, which may be practically true but fails to capture the intended semantics of any spec that semantically returns responses in, for example, $\mathbb{R}$. Furthermore, any construction that requires an oracle to
iterate through an enumerable (but potentially infinite) number of responses is absurd in practice. 

Problems in applying the SCR in practice are not restricted to the proof: the SCR statement itself may be unsuitable for conveying concrete information to implementers about how to design scalable systems in practice.
For one, the SCR is a rule that applies only for a particular commutative region, rather than all commutative regions of a spec. If a spec has 100 commutative regions, then we know each commutative region has a implementation scalable \emph{for that region}; however, we do not know if all 100 implementations are different, or if one implementation exists that will scale for all (or even multiple) commutative regions.

Furthermore, as illustrated by our oracle-based proof construction, scalability may not always be optimal: an implementation may scale but not necessarily be more performant.

In its current form, the SCR serves best as a hint that certain implementations can be made more scalable and a suggestion for areas for potential implementation optimization. For example, this use of the SCR served Clements et al. well in constructing sv6~\cite{SCR}.
It remains an open question about whether there is a way to extend the SCR, or modify the proof construction to aid implementers in designing practical, scalable systems.
Perhaps there is a way to formulate the SCR to apply not only to one particular commutative region, but rather a class of commutative regions. Or perhaps a proof construction exists for a more restricted class of commutative systems that lends itself toward efficient and practical systems.

Many other questions remain, such as whether there are specs for which all practical implementations of the spec will not commute for the SIM-commutative regions of the spec. In other words, are there specs for which the only fully scalable implementation \emph{must} have the equivalent of an oracle?
We see the space between the commutativity of implementations of a spec, and the commutativity of the spec itself as a fruitful area to
explore in future work.

\section{Acknowledgments}

This research was supported by NSF award CNS-1302359.

\bibliographystyle{abbrv}
\bibliography{n-str,n,n-conf}

\appendix
\begin{figure*}[ht!]
\begin{small}
\input{code/def}
\end{small}
\caption{Definitions for proving the SCR}
\label{fig:scrdefns}
\end{figure*}

\begin{figure*}[ht!]
\begin{small}
\input{code/thm}
\end{small}
\caption{Theorems proven about the SCR in Coq}
\label{fig:scrthms}
\end{figure*}

\end{document}